# A Ternary Digital to Analog Converter with High Power Output and 170-dB Dynamic Range

Guido Stolfi

*Abstract* – A prototype of a very high dynamic range 32-bits Digital to Analog Converter (DAC) was designed and built for the purpose of direct auditory stimulus generation. It provides signals from less than 100 nV up to 50 Watts peak power output, driving a 32-Ohms earphone or speaker. The use of ternary cells makes possible a 170 dB dynamic range that is basically limited by thermal noise only.

## I. INTRODUCTION

Audiometry test equipments should be able to generate signals from the threshold of audibility up to the discomfort level, meaning a 120-dB power range. Actually, other factors require a significantly higher span:

1. Typical audiometry headphones generate a 110 dB Sound Pressure Level (SPL) for a 0 dBm input [3]. Considering that normal auditory threshold can reach -10 $dB_{SPL}$ at frequencies near 4 kHz [1], a system noise level below -120 dBm is desirable.

2. The frequency response of the headphone can deviate up to 15 dB, so an input power of +25 dBm may be required to reach an output of 120 $dB_{SPL}$ (typical discomfort threshold) across the frequency band.

3. Short duration stimuli may require additional 10 to 20 dB to maintain perceived loudness [2].

4. Finally, another 5 dB may be required to accommodate higher quality (but lower efficiency) earphones, thus reaching a 160-170 dB dynamic range requirement.

The use of a Digital Signal Processor (DSP) for direct digital synthesis of acoustic stimulus is limited by the dynamic range of commonly available Digital to Analog Converters (DAC's), which hardly attain 130 dB even for the highest-quality Sigma-Delta devices [4]. Power amplifiers are generally needed to feed the earphones or speakers, introducing additional noise.

In this paper we describe a new DAC topology that is capable of high power output while presenting a very low quiescent noise. A prototype with 64 kHz sampling rate, allowing for flat response up to 20 kHz, was built and proven capable of generating signals from less than 100 nV to more than 50 W peak when directly coupled to a 32-Ohms earphone. Residual noise floor is totally inaudible. A 60 to 70 dB spurious-free dynamic range (SFDR) was measured over most of the signal levels, except at very low amplitudes.

## II. DAC TOPOLOGIES

Fig. 1 shows the popular R–2R ladder used in many DACs. This output impedance of this circuit is R, and the open-circuit output range approaches 0 to +V. Some drawbacks of this simple topology are:

1- Midpoint voltage is +V/2; when the generated signal has zero amplitude, the internal noise of the reference voltage (+V) is coupled to the output.

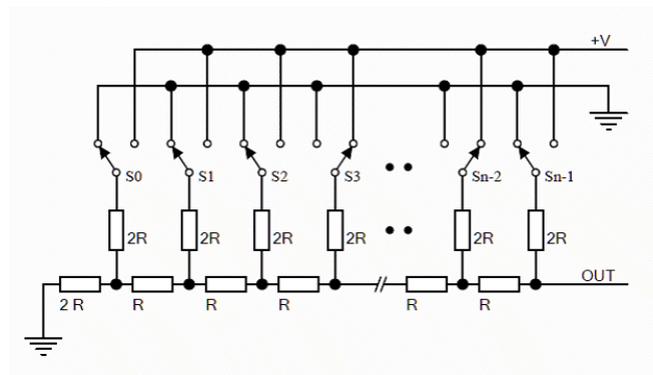

Fig. 1. Conventional R–2R DAC

2- Around midpoint, all switches toggle at the same time, introducing discontinuities at the output.

These drawbacks impair the maximum resolution of this topology, restricting its use mainly to low-cost, low resolution DAC's.

A topology which mitigates these problems is the ternary 4R-3R ladder [5], shown in Fig. 2. This circuit uses triple-throw switches connecting the ladder elements to voltages of +V, -V or Ground. Output impedance is 2R. Most remarkable features are:

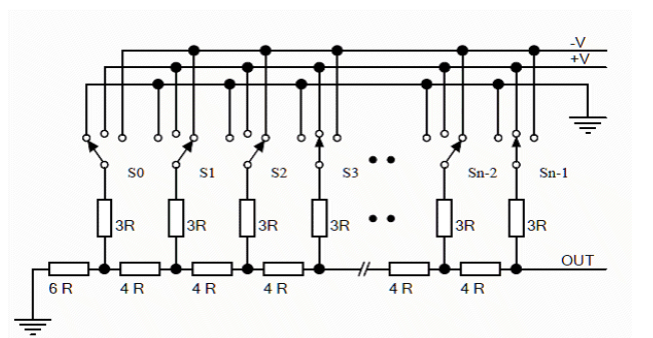

Fig. 2. Ternary (4R–3R) Ladder

The author is with the Polytechnic School, University of São Paulo – Brazil, Telecommunications Department.

1- The output signal range is centered on Ground, with maximum open-circuit amplitude approaching 2×V;

2- Every stage is controlled by a ternary digit obtained from the numeric representation of the desired output signal. A 10-stage ternary DAC will have $3^{10}$ = 59049 levels, slightly less than a 16-bit (65536 levels) binary DAC.

3- References +V and –V should be carefully matched to avoid distortion between positive and negative digits;

4- Whenever the peak-to-peak output amplitude is less than $2\times V/3^n$, all *n* switches starting from the left are permanently set to Ground; so the corresponding stages behave like a cascade of fixed attenuators, nearly 10 dB each.

This last feature is very important since all noise generated by the lower order stages, such as glitches, reference, quantization and injection noises, will be attenuated by the higher order sections. We can design a DAC with virtually unlimited dynamic range; the lower limit will be thermal noise generated by the output impedance. On the other side, we can use high voltages and low-resistance switches to get high output power without requiring an amplifier at the output.

However, triple-throw electronic switches are difficult to implement. Single-pole dual-throw (SPDT) switches are more common; for example, half-bridge MOSFET drivers, widely used in electronic power conversion, allow high voltages and currents, high speed and low ON resistance.

### III. SINGLE-SUPPLY, DIFFERENTIAL LADDER

In the proposed topology, shown in Fig. 3, the ternary ladder is split in two halves. The load, which in the present application can be speaker or a headphone, is connected differentially to both ladders.

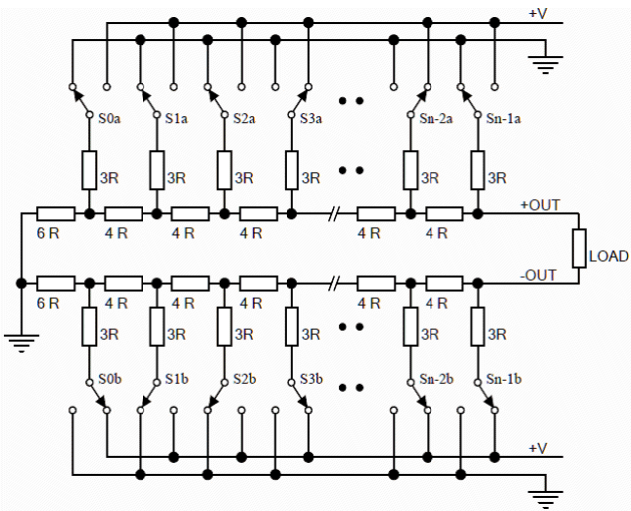

Fig. 3. Differential (4R-3R) Ternary Ladder

Positive digits are converted by the upper half and negative digits by the lower half. The same +V reference supply is used in both sides, avoiding asymmetric distortion; all switches are SPDT, and the quiescent power is zero when all switches are set to ground.

To obtain higher efficiency we can use power-of-3 weighted resistors (R – 3R – 9R – 27R…) for the most significant stages, leading to higher values (and lower dissipation) for the remaining stages. An example of this configuration is shown in Fig. 4 (only for the upper half), where the total (differential) output impedance is 1.333 R.

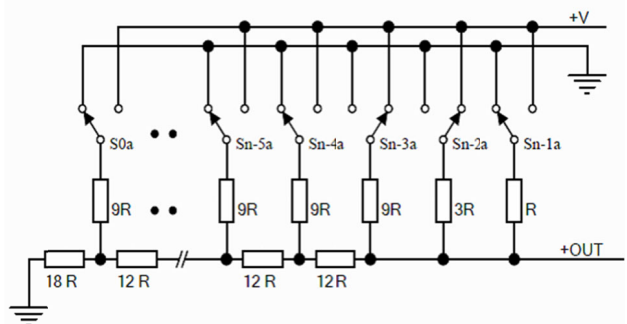

Fig. 4. Power-of-3 weighted stages (half ladder shown)

### IV. PROTOTYPE

A 20-stage prototype with 15 Ohms output impedance, based on the topology of Fig. 4, was built and tested, providing reasonable efficiency when driving dynamic earphones or speakers.

The open-circuit full scale output voltage is 180 Volts peak-to-peak. The circuit can develop 120 Vpp (42 Vrms) when connected to a 32-Ohms load, corresponding to 55 Watts RMS or +47.4 dBm. Thermal noise power at the output [6] can be computed as $Nt = kTB = 83 \times 10^{-18}$ W for a 20-kHz bandwidth (-131 dBm). The noise limited dynamic range is 178.4 dB.

To test the prototype, shown in Fig. 5, a DSP controller board was adapted to generate test signals like continuous and burst sinusoids of varying amplitudes which, after conversion to ternary representation, are transmitted to the DAC prototype by two synchronous serial ports.

Two kinds of circuit boards are used in the prototype: high-voltage modules and low-voltage / interface modules. Each high-voltage module (shown in Fig. 6) contain 4 sections: one high-current section, with 3 x 100 Ohm resistor strings in parallel, used for stages 1 and 2 of the DAC; and 3 other sections which can be programmed by jumpers to implement power-of-3 weighted resistor strings (100, 300, 900 and 2700 Ohms, used for stages 3 to 6) or 4R – 3R ladders (6000 – 8000 Ohms, used for stages 7 to 12). Eight modules are used, 4 for each half of the ladder; the most significant stage uses 3 high-current sections in parallel.

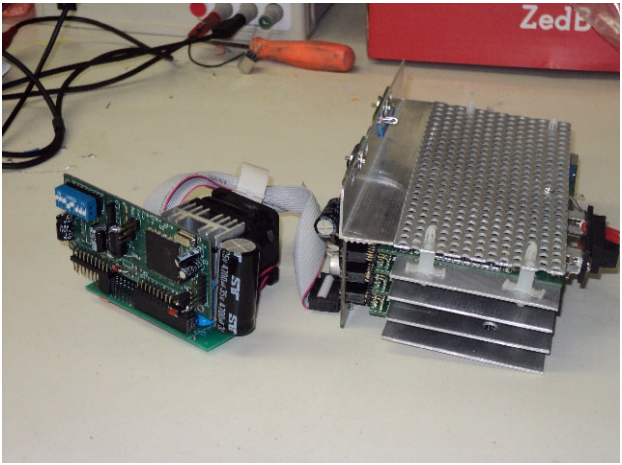

Fig. 5. DAC Prototype (right) and DSP (left)

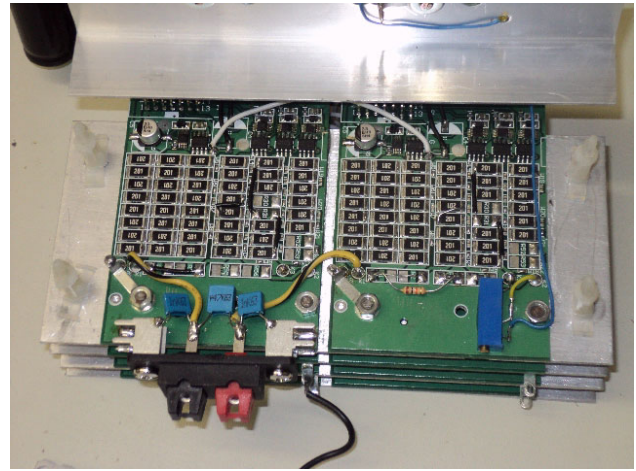

Fig. 6. High Voltage Modules

The switches are implemented with half-bridge drivers and low resistance MOSFET's, operating from a 90 Volts supply. Since the prototype was not intended to continuously generate maximum power level (only short duration bursts or clicks), it was built with low-wattage resistors and small area heat sinks.

The low-voltage / interface module (Fig. 7) contains a serial-to-parallel converter which receives the encoded digits from the DSP controller. 16 bits are sent to the high-voltage modules (4 bits per module), and 8 bits are connected to a low voltage (12 Volts), low power ladder. Two modules are used, one for the positive half and other for the negative half.

At very low levels, the electromagnetic coupling between high voltage switches and the output load (by stray capacitance, ground loops or direct radiation) would be significant. So, these last 8 stages operate with 12 Volts and use 15k - 20k Ohms resistors. Additionally, the high-voltage modules are shielded by aluminum sheets that also aid as heat sinks.

In total, there are then 20 ternary stages: 6 high-voltage power-of-3 weighted stages, 6 high-voltage ladders and 8 low-voltage ladders.

In this prototype, all signals are generated by the DSP controller as 32-bits fixed point numbers. These samples are converted to 20 ternary digits by a sequential algorithm. The resolution of this DAC is 1 to $3^{20} \approx 3.48 \times 10^9$, resulting in a theoretical maximum dynamic range of 190.8 dB due to the quantization noise. In this way, signals below the noise floor can be generated.

V. MEASUREMENTS

Fig. 8 shows the power efficiency of the prototype, measured with 1-second bursts of an 800 Hz sinusoidal tone. Maximum power supply current consumption occurs at -10 dB relative to maximum output power.

Fig. 9 shows the SFDR (Spurious-Free Dynamic Range, measured as the difference between the desired sinusoidal signal level and the highest amplitude spurious signal), with data obtained by a HP 3561A Analyzer. Measurements above +30 dBm were taken in burst mode; below -50 dBm, a low-noise preamplifier was used. It should be noted that most sections were built with regular 5% tolerance resistors. Obviously this impairs the distortion (and hence the SFDR) of the prototype. The obtained values are adequate for the intended application, being 10 to 30 dB better than typical headphone harmonic distortion [3].

Using the signal averaging function in a digital oscilloscope, the prototype operation can be verified down to the nano-Volt range. Fig.10 shows the output of the fourth low-voltage section (top trace) and the averaged DAC output after preamplifier (bottom trace), for a 60 nV burst signal. This level is -177 dB relative to full power output.

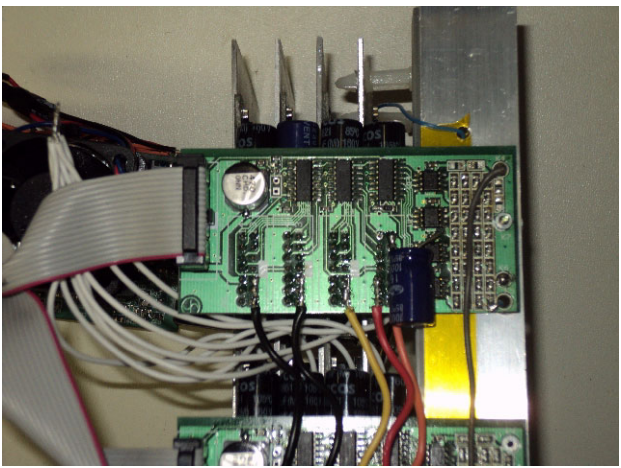

Fig. 7. Low Voltage and Interface Module

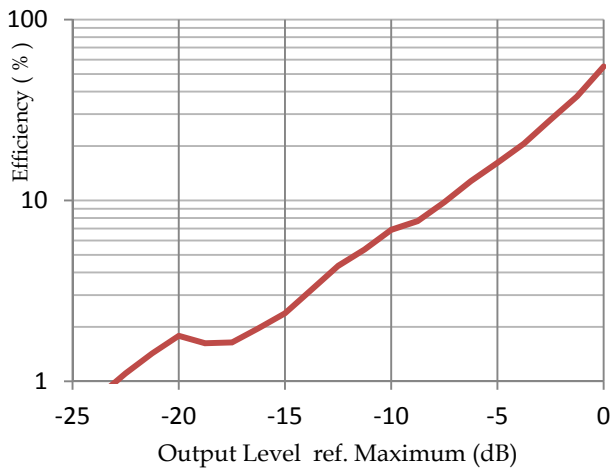

Fig. 8. DAC Power Efficiency (%)

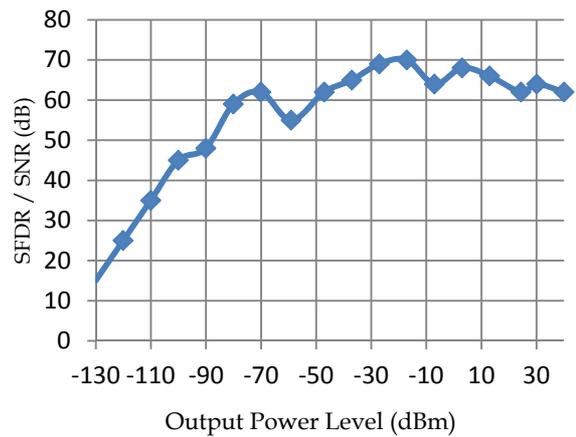

Fig. 9. DAC Spurious-Free Dynamic Range

## I. Conclusions

A prototype of a high power, low noise DAC was designed and tested, providing a seamless signal range of more than 170 dB to a speaker or headphone load. This development was originally intended for audiometry test equipment, but can be used in other applications whenever an analog signal of very high dynamic range must be generated.

## Acknowledgment

The author would like to thank Dr. L. A. Baccalá and Dr. L. A. B. Coelho for the valuable comments to this paper.

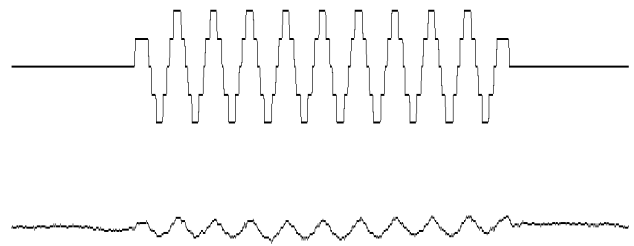

Fig. 10. Fourth low-voltage section output (top trace) and averaged DAC output (bottom trace) at 60 nV$_{RMS}$ output level